# Intense sub-2 cycle infrared pulse generation via phase mismatched cascaded nonlinear interaction in DAST crystal

A. Trisorio[1*], M. Divall[1], and C.P. Hauri[1,2]

[1]Paul Scherrer Institut, SwissFEL, 5232 Villigen PSI, Switzerland
[2]Ecole Polytechnique Federale de Lausanne, 1015 Lausanne, Switzerland
*Corresponding author: alexandre.trisorio@psi.ch

Octave-spanning, 12.5 fs, (1.9 cycle) pulses with 115 µJ energy in the short-wavelength mid-infrared spectral range (1-2.5 µm) have been generated via phase-mismatched cascaded nonlinear frequency conversion using organic DAST crystal. Such ultra-fast cascading effect is ensured by the interaction of a pump pulse with the exceptionally large effective nonlinearity of the DAST crystal and experiencing non-resonant, strongly phase mismatched, Kerr-like negative nonlinearity.

Energetic few-cycle laser pulses in the short wavelength infrared (SWIR, 1-3 µm) range are important for numerous fields of research, ranging from high harmonic generation (HHG) [1], manipulation of ultra-fast vibrational modes [2] to solar cell inspection and imaging applications. Another application of rising importance is the generation of single-cycle pulses in the terahertz (THz) spectral region (1-20 THz, so-called THz gap) via optical rectification in organic crystals [3-6]. For such crystals, ultra-broadband THz pulse generation requires a pump with ultrashort pulses located at 1.2-1.6 µm wavelength to achieve efficient energy conversion. Since the bandwidth of these THz pulses is intrinsically linked to that of the pump pulse intense and broadband SWIR pulses are an urgent need.

Over the last years various techniques have been studied in order to produce intense and ultrashort SWIR pulses at the microjoule level. Based on spectral broadening by self-phase modulation of SWIR pulses in a gas-filled hollow core fiber pulses of 400 µJ and 13.1 fs FWHM could be produced [7]. In another work, intense and self-compressed pulses were achieved by filamentation in xenon which results in 270µJ, 17.8 fs FWHM pulses [8] and even shorter ones [19]. In a different approach, four-wave mixing in a filament between the fundamental ($\omega_1$) and its second harmonic ($2\omega_1$) of a Ti:sapphire laser yield 1.5µJ, 13 fs FWHM pulses [9] ($\omega_1+\omega_1-2\omega_1\rightarrow\omega_2$). Although the nonlinear susceptibility of the gas media is small, broadband phase matching with small dispersion could be achieved. For filaments, however, the pulse energy is limited due to the critical power [10] which hinders upscaling to higher energies. Finally, we mention few-cycle pulse generation via cross-polarized wave (XPW) generation in nonlinear crystals (eg. $CaF_2$, $BaF_2$) [11] which resulted in sub-20 fs FWHM pulses. Again, the spectrally broadened and temporally cleaned pulses are limited to a few uJ pulse energy due to the intrinsic properties of the XPW effect [12].

In this Letter we experimentally demonstrate another approach to generate intense few-cycle pulses in the SWIR. The approach is based on ultra-fast (i.e non-resonant), phase-mismatched, cascaded frequency conversion in the highly nonlinear organic DAST [13] (4-N, N-dimethylamino-4'-N'-methylstilbazolium tosylate) crystal [14]. While other nonlinear crystals have been used in the past to study this process [14-16] we show that intense 2-cycle pulses are generated by using DAST, which exhibits an exceptionally large effective nonlinearity at 1600 nm ($d_{eff}\approx250$ pm/V [13]). This value is almost an order of magnitude higher than the one of the dielectric or semiconductor crystals previously studied in the literature [10]. In our experiment, the nonlinear interaction leads to a cascading of $\chi^2(2\omega;\omega;\omega)$ and $\chi^2(\omega;2\omega;-\omega)$ processes which results in octave-spanning spectral broadening (i.e. 1.2-2.5 µm) and pulse durations down to 12.5 fs (8.9 fs transform-limit) with 115µJ energy at 1850 nm. Thanks to the absence of phase matching the process can profit plainly from the large effective nonlinearity of the DAST over the entire propagation length. Moreover, the involved self-defocusing Kerr-like nonlinearity allows the technique to be up-scalable to millijoule energies by using large aperture crystals. The setup is compact and the process takes place in a noncritical interaction geometry. Finally, solitonic pulses can also be produced under specific conditions [15]. The numerical calculations support our experimental observation and our statement on the underlying involved physical effect.

Octave-spanning, self-defocusing cascaded nonlinear frequency conversion can be obtained in strongly phase mismatched condition between the fundamental wave (FW) and its second harmonic (SH), like it is the case in our experiment ($\Delta k\approx 1750$ mm$^{-1}$). During this process, the $\chi^2(2\omega;\omega;\omega)$ nonlinearity allow the FW ($\omega$) to be converted into its SH ($2\omega$). The absence of phase matching implies that after a coherence length only weak SH conversion occurs. It is then followed by a back conversion via $\chi^2(\omega;2\omega;-\omega)$ into the FW after another coherence length. During the propagation in the crystal, this process repeats itself and cascaded frequency conversion occurs. Moreover, the FW experiences nonlinear phase shift - similarly to a Kerr-like nonlinear refractive index change- , which is proportional to its intensity I: $\Delta n=n_{casc}I$. When the process is strongly phase-mismatched the refractive index can be approximated as: $n_{casc}\approx-2\omega_1 d^2_{eff}/c^2\varepsilon_0 n^2_1\ n_2 \Delta k$ [16], where $\Delta k=k_2-2k_1$ and k are the wave vectors of the fundamental and second harmonic waves.

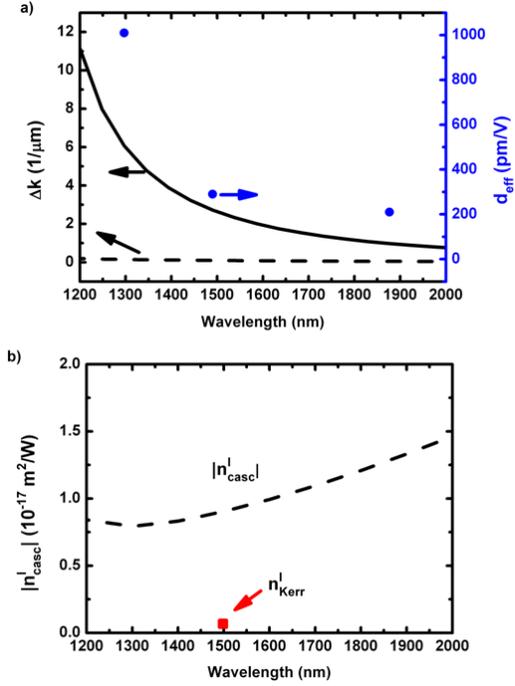

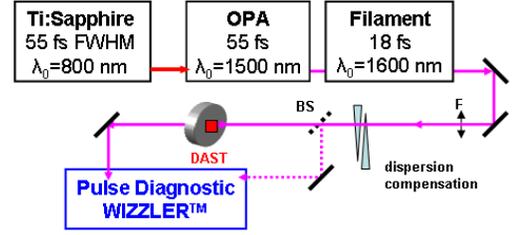

Figure 2: Experimental setup. Sub-20 fs FWHM pulses generated via an OPA and filamentation in a Xenon filled gas cell are used to generate octave spanning spectrum by phase-mismatched cascaded nonlinear interaction in DAST crystal. All pulses are characterized spectrally and temporally with a WIZZLER™ apparatus.

Figure 1: a) Calculated wavelength dependence of $\Delta k$ coming from material dispersion (black solid line), $\Delta k_r$ threshold for non resonant cascaded nonlinearity (black dashed line) and $d_{eff}$ (blue circles) in DAST. The Sellmeier formulas and values of $d_{eff}$ are taken from [13]. b) Calculated $|n^I_{casc}|$ ($n_{casc}$ <0) for DAST at 1600 nm and for $d_{eff} \approx 250$ pm/V. The red dot indicates the estimated $n_{Kerr}$ from [15].

As shown in Fig. 1 (black solid line), $\Delta k$ is approximately 1.75 μm$^{-1}$ for the pump wavelength (1600 nm). Therefore, in order to have an effective negative Kerr-like nonlinearity, $|n_{casc}|$ must be larger than $n_{Kerr}$ ($n_{Kerr}$ is the nonlinear refractive index stemming from the Kerr effect). Moreover to achieve octave-spanning frequency conversion, the cascaded nonlinearity must not be resonant. From [14], this condition is achieved provided that $\Delta k > \Delta k_r$, where $\Delta k_r = GVM_{12}/2\,GDD_2$ is a threshold for non-resonance. Here, $GVM_{12}$ is the group velocity mismatch between the FW and the SH and $GDD_2 = \partial^2 k_2 / \partial \omega^2$ is the SH group velocity dispersion. At 1600 nm, which is the pump wavelength used in our experiment, the condition $|n_{casc}| > n_{Kerr}$ is fulfilled, as shown in Fig.1 b). The values for $n_{casc}$ are computed using the Miller's scaling law [17]. As there are no dedicated studies in the literature for the measurement of $n_{Kerr}$ for the DAST an estimated value was taken from [15] for $\lambda$=1500 nm. To verify whether the off-resonance condition is fulfilled, the dependence of $\Delta k_r$ with wavelength was computed and the corresponding curve is shown in Fig. 1a) (black dashed curve). One can see that over the pump spectral range of 1200-2080 nm $\Delta k$ is effectively larger than $\Delta k_r$ insuring that off-resonance, octave spanning frequency conversion is feasible.

The experimental setup is shown in Fig. 2. A Ti:sapphire amplifier system [18] is used to drive a commercial OPA. Downstream, our post compression scheme based on filamentation produces the few-cycle pulses used to pump the DAST; a detailed description of the source can be found in [19].

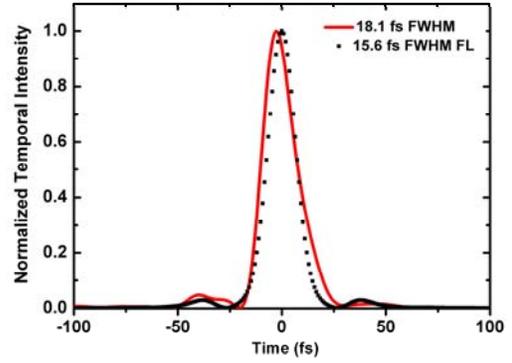

Figure 3: Single-shot reconstructed temporal profile of the pump pulse (18.1 fs FWHM - red line) at 1600 nm. The Fourier limit is 15.6 fs FWHM (black dots).

The spectral bandwidth of the pump pulse (after filamentation) covers 1450-1650 nm FWHM (fig. 4 b), blue line). The pump pulse duration is 18.1 fs FWHM, as shown in figure 3 (red line), almost Fourier limited. This corresponds to 3.1 optical cycles at 1600 nm. The 240uJ pulses are then used to pump the DAST crystal (6×6×0.18 μm$^3$) with an intensity of 6.8×10$^{10}$ W.cm$^{-2}$. In the experiment, the crystal orientation is chosen such that the pump laser polarization coincidence with the largest $d_{eff}$. This results in maximum spectral broadening. Under these conditions the spectrum undergoes significant broadening to 1350-2150 nm FWHM, as shown in Figure 4 b). An interesting feature is that the spectral broadening mainly occurs towards the longer wavelengths and seems to extend even beyond the limit set by the spectrometer used in our experiment (1100-2500 nm). The observed extensive spectral broadening is similar to previously reported numerical studies on pulse broadening via cascaded second harmonic generation in MgO:LN crystal [20]. The optical radiation emitted above 2500 nm was claimed to arise from optical Čerenkov waves that are generated simultaneously during cascaded second harmonic generation. Our experimental observation corroborates this theoretical hypothesis concerning the physical process.

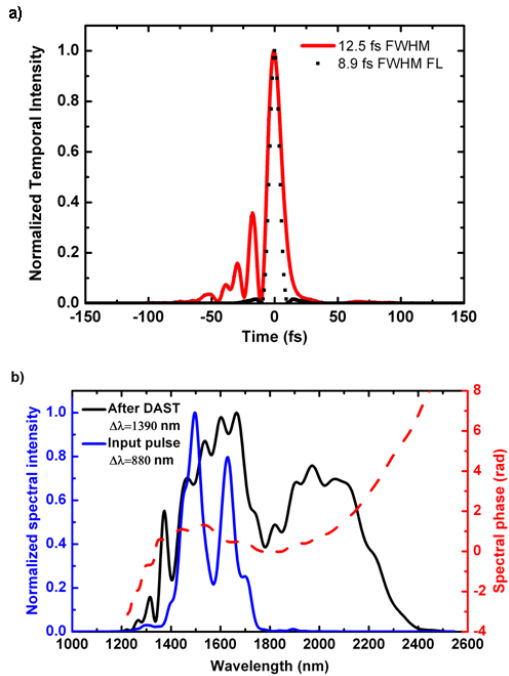

Figure 4: a) Single-shot reconstructed temporal profile of the 12.5 fs FWHM pulse (red line) after the phase-mismatched cascaded nonlinear interaction in DAST. The Fourier limit is 8.9 fs FWHM (black dots). b) Corresponding measured spectral intensity (black solid line) and spectral phase of the compressed pulse (red dashed line). As a comparison the spectrum of the pump pulse is shown in blue.

The pulse spectral bandwidth expands over 1150-2540 nm full width (Fig. 4 b), black line) and supports 8.9 fs FWHM Fourier limited pulses, as shown in figure 4 a) (black dots). This corresponds to 1.3 optical cycles at 1850 nm. The temporal pulse shape of the spectrally broadened pulse has been reconstructed using self-referenced spectral interferometry technique [18] and the results are shown in figure 4. The octave-spanning pulse exhibits a positive chirp which has been compensated by introducing a 2-mm thick fused silica plate into the beam path. After compression the pulse duration reached 12.5 fs FWHM (Fig. 4 a), red line), corresponding to 1.9 cycles at 1850 nm. However, further efforts in compression need to be undertaken to reach the transform-limited pulse duration being 8.9 fs. The observed pre-pulses shown in figure 4 are a signature of the residual third order dispersion (Fig. 4 b), red dashed line), which could not be compensated for by the simple bulk compression scheme applied in our experiment. The total pulse energy measured after compression is 115 µJ, which results in an overall efficiency of 48%. This includes all losses induced by the DAST crystal and the uncoated fused silica plate.

In conclusion, we demonstrated that phase-mismatched cascaded frequency conversion profiting from the record-large quadratic nonlinearity of the DAST crystal is well suited to generate intense, ultra-broadband few-cycle pulses in the SWIR spectral range. The simple scheme provides 1.9 cycle pulses at 1850 nm (12.5 fs FWHM), with 115 µJ energy using a DAST crystal. The extensive spectral broadening results in a pulse spectrum spanning over more than one octave (1150-2540 nm) and supports 1.3 cycle pulses (8.9 fs FWHM). Moreover, the spectrum seems to extend even further towards the mid-IR spectral range, due to Čerenkov wave generation. Our experimental results show that self-defocusing cascaded nonlinear frequency conversion technique based on DAST is a simple and efficient way to spectrally broaden SWIR pulses and to produce intense ≤2 cycle pulses. Upscaling the pulse energy towards the miliJoule level seems straightforward by use of larger crystal and a more powerful SWIR pump laser.

Acknowledgments
We acknowledge Morten Bache (DTU Fotonik-Denmark) for fruitful discussions. CPH acknowledges funding from the Swiss National Science Foundation under grant PP00P2_128493 and association to the National Center of Competence in Research (NCCR-MUST).